\documentclass[
aps,%
11pt,%
final,%
notitlepage,%
oneside,%
onecolumn,%
nobibnotes,%
nofootinbib,%
superscriptaddress,%
noshowpacs,%
centertags]%
{revtex4}

\begin{document}
\selectlanguage{english}

\title{The Structure of Clusters with Bimodal Distributions of
Galaxy Radial Velocities. II: A1775}

\author{\firstname{A.~I.}~\surname{Kopylov}}
\affiliation{\saoname}

\author{\firstname{F.~G.}~\surname{Kopylova}}
\affiliation{\saoname}

\received{March 10, 2009}%
\revised{March 24, 2009}%

\begin{abstract}
We analyze the structure of the cluster of galaxies Abell 1775
($\alpha=13^h42^m, \delta= +26\degr22'$, \mbox{$cz\approx
21000$~km/s}), which exhibits a bimodal distribution of radial
velocities of the containing galaxies. The difference of the
subcluster radial velocities is $\Delta V\approx 2900$~km/s. We
use the results of our photometric observations made with the 1-m
telescope of the Special Astrophysical Observatory of the Russian
Academy of Sciences and the spectroscopic and photometric data
from the SDSS DR6 catalog to determine independent distances to
the subclusters via three different methods: the Kormendy
relation, photometric plane, and fundamental plane. We find that
the A1775 cluster consists of two independent clusters,
\mbox{A1775A ($cz=19664$~km/s)} and A1775B ($cz=22576$~km/s), each
located at its own Hubble distance and having small peculiar
velocities. Given the velocity dispersions of 324~km/s and
581~km/s and the dynamic masses within the $R_{200}$ radius equal
to $0.6\times 10^{14}$ and $3.3\times 10^{14}$ $M_{\odot}$, the
A1775A and A1775B clusters have the K-band luminosity-to-mass
ratios of  29 and 61, respectively. A radio galaxy with an
extended tail belongs to the A1775B cluster.
\end{abstract}
\maketitle

\section{INTRODUCTION}
\vspace{0.2cm} \indent When testing the cosmological models, it is
important to know the relation between the velocities of peculiar
motions and the masses of subclusters on scale lengths exceeding
the sizes of the virialized regions of galaxy clusters. A
considerable part of rich clusters is known to consist of several
subclusters located along the same line of sight. Of special
interest are the cases where the distribution of galaxy velocities
in a cluster has a bimodal form with the difference between the
mean radial velocities of subclusters amounting to
2500--3500~km/s. Such velocity differences observed may exist
either due to the gravitational interaction between the
subclusters in massive clusters, or they may result from the
projection of unassociated clusters onto the same line of sight.
Hayashi and White~\cite{Hayashi:Kopylov_n} theoretically estimated
the limiting velocities in case of collisions of galaxy clusters
in the  $\Lambda$CDM model.

We selected four rich clusters (A1035, A1569, A1775, A1831) with a
bimodal distribution of radial velocities of the
containing galaxies ($\Delta V\sim 3000$\,km/s)
to directly (i.e., independently of their redshifts) determine the
distances to their subclusters and identify the nature of their
interaction. In the first paper of this
series~\cite{Kop1:Kopylov_n} we published the results of our study
of the  A1035 cluster and showed that the A1035A and A1035B
subclusters are actually independent clusters. The  A1775 cluster
that we study in this paper is remarkable in that it contains at
its center a close pair of giant elliptical \mbox{galaxies
\cite{Chincarini:Kopylov_n}} having a great difference between
their radial velocities, amounting to about 1900~km/s. Both
objects are radio galaxies and one of them has a very long radio
tail
\cite{Miley:Kopylov_n,Owen:Kopylov_n,Bliton:Kopylov_n,Giovan:Kopylov_n}
due to high velocity of its motion with respect to the gas of the
cluster. The presence of two subclusters in the region of A1775
was pointed out by a number of
authors~\cite{Zabludoff:Kopylov_n,Oegerle:Kopylov_n,Girardi:Kopylov_n},
however, a small number of galaxies with measured radial
velocities (about fifty) made it impossible to unambiguously
determine whether they are interacting (colliding) with each
other. It also remained unclear as to which subcluster does the
tailed radio galaxy belong. The problem of the interaction between
the subclusters is also of interest, because the A1775 cluster is
a rather powerful X-ray source, \mbox{e.g.,
\cite{Jones:Kopylov_n,Vikhlinin:Kopylov_n,Reiprich:Kopylov_n}.}
\begin{figure*}[tbp]
\setcaptionmargin{5mm}
\onelinecaptionsfalse
\includegraphics[scale=0.6,angle=-90]{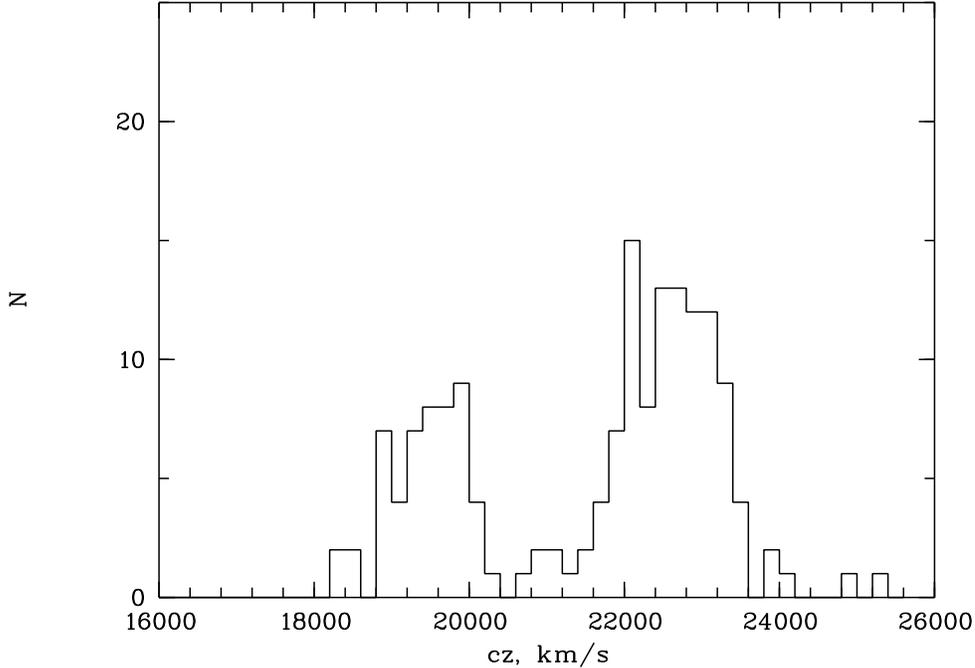}
\captionstyle{normal}
\caption{
Distribution of the radial velocities of galaxies in the region of the  A1775 cluster that are located
within  $45\arcmin$ of the double galaxy ($\alpha=13^h41^m50^s$, $\delta= +26^{0}22'19''$).}
\label{cz:Kopylov_n}
\end{figure*}

The primary aim of this paper is to determine the structure of the
A1775 cluster (the line-of-sight distance between the subclusters)
using three methods of direct distance estimates for early-type
galaxies. We use the observational material obtained with the 1-m
telescope of the Special Astrophysical Observatory of the Russian
Academy of Sciences (the SAO RAS) and the data of the SDSS DR6
catalog (Sloan Digital Sky Survey Data Release 6).  We also use
the data from the 2MASS catalog to determine the infrared
luminosities of the clusters.

The present paper has the following layout. In Section 2 we
describe the samples of early-type galaxies; in Section~3 we
determine direct distances to the subclusters of the  A1775
cluster. In Conclusions we discuss the results obtained. We adopt
here the following cosmological parameters: $\Omega_m=0.3$,
$\Omega_{\Lambda}=0.7$, \mbox{$H_0=70$~km/s/Mpc}.

\section{DESCRIPTION OF DATA}
\vspace{0.2cm} \indent In this section we describe the
observational data for early-type galaxies that we use to
determine the relative distances to the subsystems in the  A1775
cluster. According to Abell's catalog~\cite{Abell:Kopylov_n}, the
radial velocity of the cluster is \mbox{$cz\simeq 21000$~km/s,} it
has richness class of 2 and belongs to the Bautz-Morgan I type.
The data of the SDSS DR6 catalog \cite{Adelman:Kopylov_n} allow
two subclusters (A1775A and A1775B) to be identified. This becomes
evident from the distribution of radial velocities in the cluster
region, which we demonstrate in Fig.~\ref{cz:Kopylov_n}. The radio
galaxy with a long tail has a radial velocity of $cz=20812$~km/s,
which is located approximately midway between the peaks of the
bimodal distribution. The 2.5 $\sigma$-criterion of galaxy
selection, which we usually employed to select cluster members
\cite{Kop2:Kopylov_n}, marks this galaxy as a nonmember of the
A1775B subcluster. However, there is evidence suggesting that this
galaxy interacts with the brightest galaxy at the center of the
A1775B cluster. The hot gas in the cluster is substantially
perturbed~\cite{Anders:Kopylov_n}: the X-ray radiation of the gas
in the central region exhibits temperature and surface-brightness
irregularities, and the center of the X-ray radiation does not
coincide with the brightest galaxy in the cluster due to,
apparently, the high-velocity flyby of the tailed radio galaxy
across the center of the cluster. We therefore used a softer
\mbox{$3\sigma$-criterion} to include it into the A1775B
subcluster.

\begin{table*}[tbp]
\setcaptionmargin{0mm} \onelinecaptionstrue
\captionstyle{flushleft}
\caption{Cluster data}
\label{cluster:Kopylov_n}
\medskip
\begin{tabular}{l|c|c} \hline
 Cluster properties &  \quad \quad \quad A1775A \quad \quad \quad &  \quad \quad \quad A1775B \quad \quad \quad \\
\hline
$\alpha$~(J2000)      & $13^h 42^m 41.^s99$   & $13^h 41^m 49.^s14$  \\
$\delta$~(J2000)      &$ +26\degr 14' 23.''2$   &$ +26\degr 22' 24.''5 $ \\
$z_h$                          & 0.065591      & 0.075138     \\
$cz_h$, km/s           & 19664         & 22576        \\
$\sigma$, km/s          & $324\pm76$    & $581\pm74$   \\
$R_{200}$, Mpc                 & 0.78          & 1.39         \\
$N_{200}$                      & 18            & 62           \\
$M_{200}$, $10^{14}~M_{\odot}$ & $0.57\pm0.40$ & $3.28\pm1.25$\\
$L_{200}$, $10^{12}~L_{\odot}$ & $1.99\pm0.11$ & $5.34\pm0.05$\\
$M/L_K$, $M_{\odot}/L_{\odot}$ & $29\pm21$     & $61\pm24$    \\
$L_X(0.1-2.4\, {\rm keV})$, $10^{44}~{\rm erg/s}$ & $-$ & 1.6          \\
\hline
\end{tabular}
\end{table*}

\begin{figure*}[tbp]
\setcaptionmargin{5mm}
\onelinecaptionsfalse
\includegraphics[scale=0.6,angle=-90]{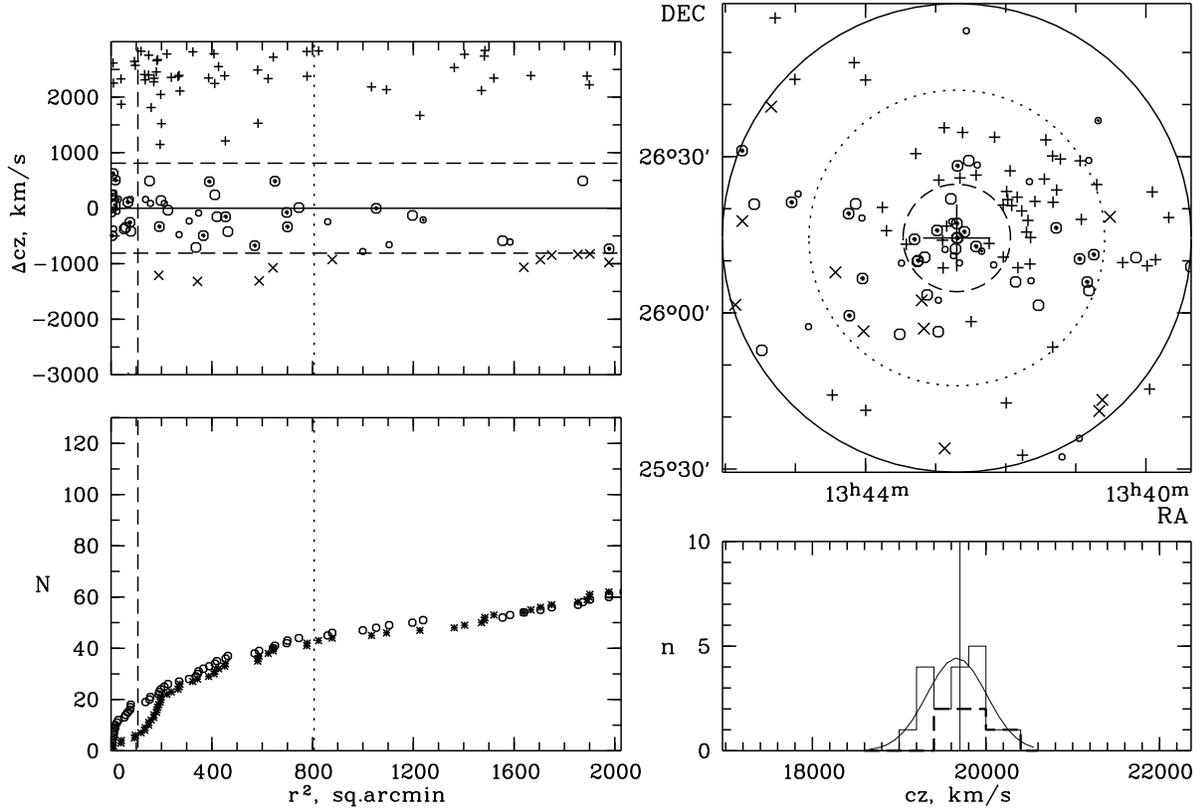}
\captionstyle{normal} \caption{ Distribution of member galaxies in
A1775A. The {\it top left} figure shows the deviation of the
radial velocities of the galaxies from the mean radial velocity of
the cluster as determined from the galaxies located inside the
$R_{200}$ radius. The horizontal dashed lines correspond to the
$\pm2.5\sigma$ deviations; the vertical dashed line indicates the
radius $R_{200}$; the \mbox{dotted line} is the Abell radius
(2.143 Mpc). Big circles indicate the galaxies that are brighter
than $M_K^*+1 = -23^m.29$; the circles with a dot at the center
show early-type galaxies; the plus signs are the background
galaxies. The {\it bottom left} panel shows the integrated
distribution of the number of galaxies as a function of the
squared angular clustercentric distance. The circles correspond to
the symbols in the top left figure and \mbox{asterisks}---to all
field galaxies. The {\it top right} figure shows the sky
distribution of the galaxies represented in the top left figure in
equatorial coordinates (the same symbols are used). The circles
indicate the $R_{200}$- (dashes) and Abell-radius (dots) domains.
The region studied is bounded by the circle of radius 45$\arcmin$
(the solid line). The large cross indicates the center of the
cluster. The {\it bottom right} figure shows the distribution of
the radial velocities of the cluster member galaxies located
within the $R_{200}$ radius (the solid line is the Gaussian
corresponding to this distribution) and of early-type galaxies
(the dashed line). The solid vertical line indicates the mean
radial velocity of the cluster. } \label{paspA:Kopylov_n}
\end{figure*}

\begin{figure*}[tbp]
\setcaptionmargin{5mm}
\onelinecaptionstrue
\includegraphics[scale=0.6,angle=-90]{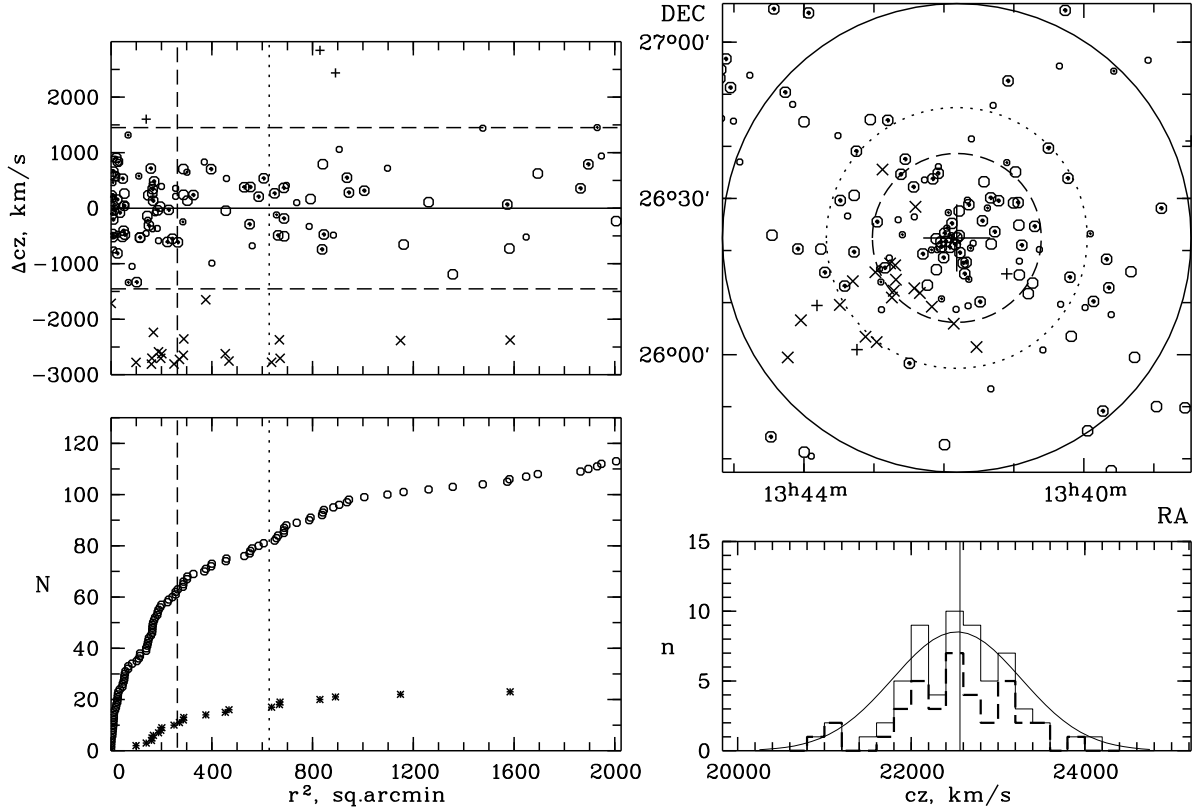}
\captionstyle{normal}
\caption{Distribution of galaxies in A1775B. The layout and designations are the same as in Fig.~2 except for
the top left panel, where the horizontal dashed lines correspond to  $\pm3\sigma$ deviations.}
\label{paspB:Kopylov_n}
\end{figure*}

Figures~\ref{paspA:Kopylov_n} and  \ref{paspB:Kopylov_n} show our
estimated values for the main cluster parameters: the deviations
of the radial velocities of the cluster member galaxies from the
mean radial velocity of the cluster; the integrated distribution
of the number of galaxies as a function of the squared
clustercentric distance (to better illustrate the identification
of the cluster core, the outer boundary of the cluster halo, and
the region dominated by a uniform environment---the linear portion
of the relation); the positions of the galaxy in the sky plane;
the distribution of radial velocities of all cluster galaxies
located within  $R_{200}$ (the corresponding Gaussian is shown)
and that of the radial velocities of  early-type galaxies inside
the same radius. The center of the A1775B cluster coincides with
the brightest  cD galaxy UGC 08669. The center of A1775A hosts the
brightest elliptical galaxy of the cluster, which is located close
to its centroid.

Table~\ref{cluster:Kopylov_n} lists our estimates for the
parameters of the clusters that we determined for the
$R_{200}$-radius region based on  SDSS data supplemented by the
radial velocity measurements adopted from the NED database. Here
$R_{200}$ is the radius of the virialized part of the cluster
where the mass density is 200 times higher than the critical
density of the Universe. The cluster mass inside the region
bounded by this radius is determined by the dispersion of galaxy
velocities~\cite{Kop2:Kopylov_n}, which we list in the Table (with
the cosmological correction $(1+z)^{-1}$ applied). The error of
the inferred mass of the cluster is determined by the error of
velocity dispersion. Moreover, the table also lists the total
luminosities of the clusters computed using galaxies with near-IR
luminosities lower than $M_{K,lim} = -21^m$, and the ratio of the
inferred mass-to-IR luminosity. See~\cite{Kop2:Kopylov_n} for a
description of the technique used to estimate the luminosities.
The mass-to-IR luminosity ratio for A1775A and A1775B does not
differ significantly from the ratio that we earlier
obtained~\cite{Kop2:Kopylov_n} for a large sample of clusters of
galaxies. We adopt the  \mbox{0.1--2.4\,keV} luminosity
from~\cite{Reiprich:Kopylov_n}. Our estimate for the mass of the
A1775B cluster agrees well with the mass measured by~Reiprich and
B\"{o}hringer~\cite{Reiprich:Kopylov_n} by the \mbox{X-ray}
luminosity, $M_{200} = 4.22^{+0.59}_{-0.40} \times
10^{14}~M_{\odot}$.

\begin{table*}[tbp]
\setcaptionmargin{0mm} \onelinecaptionstrue
\captionstyle{flushleft}
\caption{Parameters of early-type galaxies in  A1775 as measured from  $R_c$-band images taken with the
1-m telescope}
\label{data:Kopylov_n}
\medskip
\begin{tabular}{c|c|c|c|c|c|c|c} \hline
Cluster& $\alpha~~(J2000)~~\delta$& $z_h$& $cz_h$& $m_R$& $R_e$&     $\mu_e$& $n$\\
\hline
       & hhmmss ddmmss          &      & km/s& mag. & arcsec & mag./$\sq''$&    \\
\hline
A1775A & 13 42 42.01+26 14 23.4 & 0.065362 & 19595 & 14.01 & 10.30 & 22.39 & $5.12\pm0.49$\\
 & 13 42 25.61+26 12 44.7 & 0.065767 & 19716 & 14.68 &  3.36 & 20.44 & $2.56\pm0.17$\\
 & 13 42 59.01+26 15 49.3 & 0.067281 & 20170 & 14.70 &  4.21 & 20.95 & $3.25\pm0.37$\\
 & 13 43 15.28+26 10 02.4 & 0.064738 & 19408 & 14.87 &  4.17 & 21.14 & $3.14\pm0.35$\\
 & 13 40 57.06+26 10 21.5 & 0.063339 & 18989 & 15.77 &  2.73 & 20.91 & $1.02\pm0.08$\\
 & 13 43 18.32+26 14 07.1 & 0.065951 & 19772 & 15.81 &  2.98 & 21.21 & $2.84\pm0.38$\\
 & 13 40 44.89+26 11 11.1 & 0.064471 & 19328 & 15.84 &  2.94 & 21.21 & $1.59\pm0.22$\\
& 13 43 15.84+26 09 52.8 & 0.064203 & 19248 & 16.79 &  2.11 & 21.04 & $1.48\pm0.99$\\
\hline
A1775B & 13 41 49.14+26 22 24.5 & 0.075732 & 22704 & 13.48 & 27.24 & 23.72 & $4.78\pm0.33$\\
 & 13 41 50.46+26 22 13.0 & 0.069420 & 20812 & 14.25 &  9.26 & 22.18 & $4.35\pm0.41$\\
 & 13 40 56.59+26 29 12.2 & 0.075010 & 22487 & 14.68 &  6.75 & 22.09 & $2.99\pm0.25$\\
 & 13 42 02.84+26 21 38.2 & 0.075253 & 22560 & 15.36 &  4.23 & 21.73 & $4.24\pm0.81$\\
 & 13 42 05.13+26 34 49.3 & 0.075582 & 22659 & 15.66 &  2.61 & 20.73 & $2.32\pm0.38$\\
 & 13 41 55.13+26 20 35.7 & 0.073588 & 22061 & 15.99 &  2.13 & 20.45 & $2.01\pm0.25$\\
 & 13 42 18.28+26 19 20.2 & 0.075282 & 22569 & 16.15 &  2.61 & 21.23 & $2.08\pm0.30$\\
 & 13 41 50.60+26 21 10.6 & 0.075899 & 22754 & 16.17 &  2.39 & 21.04 & $1.54\pm0.20$\\
 & 13 42 09.78+26 33 44.8 & 0.074096 & 22213 & 16.25 &  2.42 & 21.09 & $2.26\pm0.49$\\
 & 13 43 17.28+26 19 43.2 & 0.077482 & 23229 & 16.30 &  2.61 & 21.35 & $1.18\pm0.17$\\
 & 13 42 02.46+26 20 43.2 & 0.075025 & 22492 & 16.44 &  2.04 & 20.83 & $1.71\pm0.43$\\
\hline
\end{tabular}
\vspace{0.5 cm}
\end{table*}

\subsection{Parameters of Early-Type Galaxies as Measured with the 1-m Telescope of the SAO RAS}
We determined the photometric parameters of 19 galaxies in the
subclusters studied by analyzing direct $R_c$-band (Cron--Cousins
system) images that we obtained using the 1-m telescope of the SAO
RAS in April 1999. These images were obtained under intermediate
seeing conditions of $1.\arcsec65 $ (the FWHM of stellar-image
profiles). We used a $520\times580$ ISD015A CCD with pixel size of
$18\times24$ $\mu$m$^2$ corresponding to an angular size of $(0.28
\times 0.37)\,\square \arcsec$. The exposure time was equal to
500~s. We observed Landolt standard stars~\cite{Landolt:Kopylov_n}
several times during the night in order to establish the
photometric calibration.

We used the MIDAS software package (Munich Image Data Analysis
System) to reduce the observational data. We adopt a standard
procedure of image reduction: subtraction of median dark frame,
division by a flat field, and subtraction of sky background
approximated by a second-order surface. We then determine the
asymptotic total magnitude of the galaxy via multiaperture
photometry and use this magnitude to compute the effective radius
$R_e$, containing half of the total luminosity of the galaxy, and
the effective surface brightness $\mu_e$ at this radius. We
determine the parameter  $n$ characterizing the shape of the
surface-brightness profile by fitting the Sersic
profile~\cite{Sersic:Kopylov_n} $R^{1/n}$ ($n=4$ for the \mbox{de
Vaucouleurs \cite{deV:Kopylov_n}} profile) to the observed profile
in the galactocentric radius interval from \mbox{$3\times $FWHM}
out to the radius where the surface brightness is equal to
\mbox{(24--25) mag./$\square''$.} To determine the distances, we
then corrected the resulting photometric parameters of galaxies
$R_e$, $\mu_e$ for seeing effects using the method described by
\mbox{Saglia et al.~\cite{Saglia:Kopylov_n}}. This correction is
equal to $19\%$ for the galaxies with the effective radii smaller
than 3$\arcsec$. Hence, in this work we use model-independent
galaxy parameters ($R_e$, $\mu_e$) estimated by the total
asymptotic magnitude and the model dependent \linebreak value $n$.

We list the results of our photometric measurements in
Table~\ref{data:Kopylov_n}. It gives the following observed
(seeing corrected) parameters of galaxies: the number of the
cluster according to the Abell catalog \cite{Abell:Kopylov_n}; its
J2000 equatorial coordinates; heliocentric redshift and radial
velocity (according to SDSS or NED); the total (asymptotic)
magnitude; effective radius in arcsec; effective surface
brightness at the effective radius, and parameter $n$ of the
Sersic profile with its error. We determined the observed
parameters for the pair of the brightest giant galaxies in the
A1775B cluster by fitting the observed profiles to the Sersic
profile.

\subsection{Parameters of Early-Type Galaxies According to the SDSS Catalog}
We compiled a sample of early-type galaxies in the A1775A and
A1775B clusters based on the data of the SDSS (DR6) catalog
($r$-band filter). We selected the galaxies in accordance with the
criteria described by Bernardi et al.~\cite{Bern3:Kopylov_n} down
to an extinction-corrected apparent Petrosian magnitude of
$17^m.77$. We supplemented these criteria by the concentration
index of the galaxy in this filter. To reduce the effect of the
errors while determining the stellar velocity dispersion in a
galaxy, we only selected the objects with $\sigma$ greater than
100~km/s. Neither did we not use galaxies with the effective radii
smaller than 1$''$, as no such galaxies were found in A1775A,
whereas the errors of the galaxy parameters determination increase
with the decreasing size of the galaxy. We found a total of 21 and
32 galaxies inside the virialized regions of the clusters and
inside the Abell radius (2.143~Mpc in the adopted model),
respectively. Table~\ref{sdss:Kopylov_n} lists the following
parameters of the sample of early-type galaxies located within the
$R_{200}$ radius: the J2000 equatorial coordinates; heliocentric
redshift and radial velocity; central dispersion of stellar
velocities $\sigma$; the parameters of the de Vaucouleurs profile
(the total magnitude and the effective radius multiplied by
$\sqrt{b/a}$); \mbox{$fracDeV_r\geq 0.8$}, the quantity that
characterizes the contribution of the de Vaucouleurs bulge to the
surface brightness profile of the galaxy; $r_{90}/r_{50}\geq 2.6$,
the concentration index, which is equal to the ratio of the radii
containing  90\% and 50\% of the Petrosian flux; $eClass\leq 0$,
the parameter characterizing the spectrum of the galaxy: minus
means that the spectrum exhibits no appreciable emission lines.
The SDSS catalog gives wrong parameters for the ($13^h41^m50.45^s
+26\degr22'13''$) galaxy and we therefore used our data (Table\,2)
for the effective radius and the  $r$-band magnitude to compute
the average surface brightness (Section 3):
 \mbox{$r=R_с+0.2936\times(r-i)+0.1439$ \cite{Lupton:Kopylov_n},}
for the typical early-type galaxy color of \mbox{$r-i = 0^m.41$.}

\begin{table*}[tbp]
\setcaptionmargin{0mm} \onelinecaptionstrue
\captionstyle{flushleft}
\caption{Parameters of early-type galaxies in  A1775 according to the SDSS catalog}
\label{sdss:Kopylov_n}
\medskip
\begin{tabular}{c|c|c|c|c|c|c|c|c|c}
\hline
 Cluster& $\alpha~~(J2000)~~\delta$& $z_h$&     $cz_h$&    $\sigma$& $m_r$&
 $R_e$& $fracDeV_r$& $r90/r50$& $eClass$ \\
\hline
       & hhmmss ddmmss        &      &km/s& km/s& mag.&   arcsec &            &          &          \\
\hline
A1775A& 13 42 25.61+26 12 44.7& 0.065767& 19716& 241& 14.914&  3.146& 1.00& 3.34& -0.121\\
      & 13 42 59.01+26 15 49.3& 0.067281& 20170& 236& 14.960&  3.861& 1.00& 3.34& -0.125\\
      & 13 42 20.84+26 11 50.5& 0.066127& 19824& 104& 17.089&  1.665& 0.96& 2.83& -0.124\\
      & 13 42 42.02+26 17 09.7& 0.067674& 20288& 210& 15.882&  1.512& 0.93& 3.16& -0.135\\
      & 13 43 18.32+26 14 07.1& 0.065951& 19772& 163& 15.939&  3.086& 1.00& 3.28& -0.133\\
      & 13 42 41.39+26 14 23.2& 0.066392& 19904& 119& 17.206&  1.219& 0.98& 3.93& -0.091\\

\hline
A1775B& 13 41 50.45+26 22 13.0& 0.069420& 20812& 300& 14.510&  9.260&    -&    -& -0.143\\
      & 13 42 05.13+26 34 49.3& 0.075582& 22659& 191& 15.719&  2.364& 1.00& 3.11& -0.141\\
      & 13 42 26.23+26 32 12.6& 0.076281& 22868& 112& 15.984&  5.828& 1.00& 2.96& -0.119\\
      & 13 42 18.28+26 19 20.2& 0.075282& 22569& 165& 16.329&  2.411& 1.00& 3.19& -0.136\\
      & 13 42 09.78+26 33 44.8& 0.074096& 22213& 149& 16.349&  2.230& 1.00& 2.88& -0.128\\
      & 13 42 02.46+26 20 43.2& 0.075025& 22492& 170& 16.584&  1.655& 1.00& 3.02& -0.143\\
      & 13 41 38.90+26 28 47.3& 0.076912& 23058& 172& 16.665&  1.197& 0.94& 2.69& -0.141\\
      & 13 42 50.81+26 16 40.5& 0.073104& 21916& 155& 16.677&  1.456& 1.00& 3.29& -0.120\\
      & 13 41 13.43+26 29 33.2& 0.073379& 21998& 107& 16.987&  1.509& 0.98& 2.84& -0.100\\
      & 13 42 57.45+26 25 30.0& 0.073298& 21974& 111& 16.698&  2.080& 1.00& 2.98& -0.112\\
      & 13 42 02.30+26 10 42.7& 0.074409& 22307& 102& 17.215&  1.598& 1.00& 2.84& -0.113\\
      & 13 42 00.48+26 18 38.1& 0.078156& 23431& 135& 17.273&  1.140& 0.97& 2.82& -0.130\\
      & 13 40 42.35+26 24 38.1& 0.075034& 22495& 166& 17.344&  1.369& 1.00& 3.25& -0.129\\
      & 13 41 29.16+26 10 08.1& 0.076746& 23008& 117& 17.375&  1.903& 0.84& 2.67& -0.123\\
      & 13 41 56.48+26 27 16.4& 0.073569& 22055& 110& 17.599&  1.335& 1.00& 2.70& -0.097\\
\hline
\end{tabular}
\end{table*}

\section{DETERMINATION OF THE RELATIVE DISTANCE BETWEEN THE SUBCLUSTERS IN A1775}
Redshift-independent methods of the determination of distances to
objects (in our case, to clusters of galaxies) are of fundamental
importance in cosmology. The distances to the clusters of galaxies
are often determined using the parameters of early-type galaxies,
which are the dominant population in the central regions of
clusters. Some of these parameters depend on the distance (e.g.,
radius or luminosity), whereas other parameters  are distance
independent  (surface brightness or velocity dispersion). In this
paper we use three methods, each based on a combination of these
parameters of early-type galaxies: the Kormendy relation
\cite{Kormendy:Kopylov_n}, the photometric plane \mbox{(PP)
\cite{Graham:Kopylov_n},} and the fundamental plane (FP)
\cite{Djorgovski:Kopylov_n}.

\begin{figure*}[tbp]
\setcaptionmargin{5mm}
\onelinecaptionsfalse
\includegraphics[scale=0.6,angle=-90, bb= 51 27 451 793,clip]{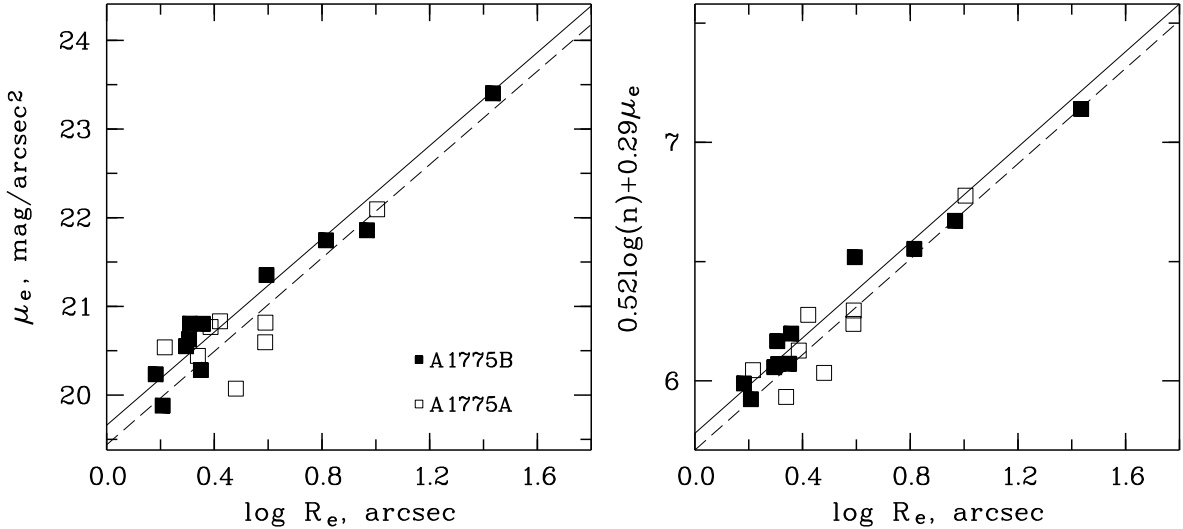}
\captionstyle{normal}
\caption{
The Kormendy relation (left) and photometric plane (right) for early-type galaxies in A1775A and A1775B
according to the data obtained with the 1-m telescope. The dashed and solid lines correspond to the
zero points of clusters A and B, respectively.}
\label{kr:Kopylov_n}
\end{figure*}

As we pointed out above, in case of A1775 (a cluster with a
bimodal distribution of radial velocities) we have two
alternatives: either the subclusters A1775A and A1775B are
gravitationally bound, located at the same distance, and make up a
single large cluster, or they are gravitationally unbound
independent clusters located at different distances. In the latter
case they should obey the Hubble law, which relates radial
velocity and distance. A detailed description of the determination
of cluster distances via the Kormendy relation
\cite{Kormendy:Kopylov_n} can be found in ~\cite{Kop3:Kopylov_n}.
This relation for a large sample of galaxies has the form:
\mbox{$\log R_e=0.38\mu_e + \gamma$.} Figure~\ref{kr:Kopylov_n}
(left) shows this relation as derived from our observations of
eight galaxies in A1775A and 11 galaxies in A1775B. We visually
selected these galaxies in 1999 on the Palomar/Las Campanas Atlas
of Nearby Galaxies images among the galaxies with early-type
morphological characteristics and with then known radial
velocities. The x-axis gives the seeing-corrected observed $\log
R_e$ in arcseconds. The y-axis gives the surface-brightness values
with the cosmological correction $10\log(1+z)$ applied. The
K-correction and the evolutionary correction at the redshifts $z$
considered are approximately equal in the absolute value, but have
opposite signs. We derive the following zero points prior to
applying the magnitude correction: $\gamma_A=-7.389~(rms=0.148)$,
$N=8$; \mbox{$\gamma_B=-7.470~(rms=0.088)$,} \mbox{$N=11$.} The
difference between these zero points is
$\gamma_{AB}=+0.081\pm0.058$. When corrected for the dependence of
the residuals from this relation on galaxy magnitudes the zero
points become $\gamma_A=-7.372~(rms=0.087)$ and
\mbox{$\gamma_B=-7.458~(rms=0.072)$.} The zero-point difference is
$\gamma_{AB}=+0.086\pm0.038$. This difference should be equal to
0.059 if the two subsystems obeyed the Hubble law.

The photometric plane can be derived from the fundamental plane
for early-type galaxies using the photometrically measured
parameter $n$ characterizing the form of the Sersic profile
instead of the spectroscopically measured parameter---the central
velocity dispersion of stars in the galaxy. A photometric plane
was derived, e.g., in the paper by Graham~\cite{Graham:Kopylov_n}.
To derive our  $R_c$-band photometric plane, we use the
photometric parameters  $R_e$ and $\mu_e$ measured by Kopylov and
Kopylova \cite{Kop4:Kopylov_n} for 12 early-type galaxies on the
images taken with the 6-m telescope of the SAO RAS under $1''$
seeing conditions. We determine the parameter $n$ from the
surface-brightness profile. These data yield the following
photometric plane: \mbox{$\log R_{e}= 0.52(\pm0.130)\log
n+0.29(\pm0.03)\mu_{e}+\gamma$}, which we show in
Fig.~\ref{kr:Kopylov_n} (right). We derived the following zero
points for the two subsystems in  A1775:
$\gamma_A=-5.712~(rms=0.108)$, $N=8$; $\gamma_B=-5.769~$
$(rms=0.083)$, $N=11$. The difference between these
zero points is  $\gamma_{AB}=+0.057\pm0.045$.

\begin{figure}[tbp]
\setcaptionmargin{5mm}
\onelinecaptionstrue
\includegraphics[scale=0.35,angle=-90]{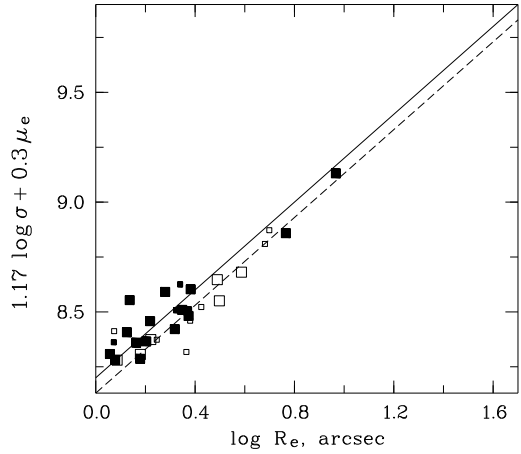}
\captionstyle{normal}
\caption{
Fundamental plane for early-type galaxies in the  A1775A and A1775B clusters (according to SDSS DR6 data)
located within the Abell radius. Larger symbols show the galaxies located within the  $R_{200}$
radius. Their parameters are listed in Table~3. }
\label{fp:Kopylov_n}
\end{figure}

The SDSS data, which are available for a greater number of
galaxies in the A1775A and A1775B clusters, allow us to more
accurately estimate the zero points  (i.e., the distances to the
clusters) using the fundamental plane, because the statistical
accuracy depends on the number of galaxies. To derive the
parameters of the fundamental plane, we compute the average
effective surface brightness by the following formula:
\mbox{$<\mu_e> = r+2.5 \log(2 \pi R_e^2)-10 \log(1+z)$.} We reduce
the central velocity dispersion $\sigma$ and the effective radius
to a circular aperture as described by \mbox{Bernardi et
al.~\cite{Bern1:Kopylov_n}.} Figure~\ref{fp:Kopylov_n} shows the
fundamental plane for the selected early-type galaxies located
inside the Abell radius. Larger symbols indicate the galaxies
located inside the $R_{200}$ radius. Direct regression on  $\log
R_e$, which \mbox{Bernardi et al. \cite{Bern2:Kopylov_n}} derived
for 9000 galaxies of the SDSS catalog, has the form \linebreak
\mbox{$\log R_{e}= 1.17\log \sigma+0.30 <\mu_e> + \gamma$.} We
derived the following zero points for the subclusters in A1775 for
the $R_{200}$ and Abell radii (in the parentheses):
$\gamma_A=-8.130~(-8.129)$ $(rms=0.050~(0.088))$, $N=6~(13)$;
$\gamma_B=-8.202~(-8.206)$  $(rms=0.090~(0.086))$, $N=15~(19)$.
The difference of the zero points is equal to
$\gamma_{AB}=+0.072\pm0.030(+0.077\pm0.031)$.
As a result, both the direct and orthogonal
regressions~\cite{Bern2:Kopylov_n} yield an $R_{200}$ value of
$\gamma_{AB}=+0.060\pm 0.021$, i.e., the distances
to the two subclusters differ by about 3$\sigma$.

Hence, all the distance-estimation methods imply, albeit with
different accuracy, that the subclusters in A1775 do not form a
single bound system separated from the Hubble flow, but are
independent clusters. We also determine the average peculiar
velocities of A1775A and A1775B corresponding to the deviations of
their $\log R_e$ from the zero points of direct and orthogonal
regression, which are equal to --8.046 and --8.763, respectively.
We determine the corresponding zero points based on the data for
423 early-type galaxies in the region of the Ursa Major
supercluster \cite{Kop5:Kopylov_n} with $\log R_e$ > 0,
\mbox{$\log \sigma$ > 100~km/s}, and \mbox{$M_r > -20^m$.} The
average peculiar velocities of the nearest and farthest clusters
(A1775A and A1775B) inferred from galaxies located within
$R_{200}$ are equal to ~\mbox{$(+751\pm504)$} and
\mbox{$(-99\pm866)$~km/s}, respectively.

\section{CONCLUSIONS}
\vspace{0.2cm} \indent The A1775 cluster has many interesting
peculiarities. Firstly, it has a bimiodal distribution of radial
velocities (Fig.~\ref{cz:Kopylov_n}), which indicates that the
cluster actually contains two subclusters. Secondly, the cluster
hosts at its center an unusually close pair of radio galaxies: one
of them is the brightest galaxy in the cluster and it is located
near the X-ray center of A1775B, whereas the velocity of the
second radio galaxy, which has a long radio tail, is intermediate
between the radial velocities of the two subclusters. We consider
this second radio galaxy to be a member of A1775B because of clear
evidence indicative of its interaction with both the brightest
galaxy and with the hot gas in the center of A1775B.

In this paper we attempted to find out whether the subclusters are
gravitationally bound to each other by estimating the distance
between these clusters. We measured the $R_c$-band photometric
parameters ($m_R$, $\mu_e$, $\log R_e$, $n$) for 19 early-type
galaxies in the subclusters A1775A and A1775B on the CCD frames
that we took with the 1-m telescope of the SAO RAS. We used these
data to derive the Kormendy relation and photometric plane for
early-type galaxies. We also used the data from the SDSS DR6
catalog to determine the principal parameters of the clusters
considered and the parameters of the $r$-band fundamental plane
for early-type galaxies. The distances to the clusters measured
using  three methods allowed us to determine the dynamic state of
A1775 and estimate the peculiar velocities of its subsystems. We
conclude that  A1775 consists of two independent subclusters
located at different Hubble distances. Note also that given their
radial-velocity difference of  2900\,km/s, our inferred virial
masses of the subclusters are too small for these groups to be
gravitationally bound even if they are located at the same
distance.

\begin{acknowledgments}
This work was supported in part by the Russian Foundation for
Basic Research (grant No 07-02-01417a). This research has made use
of the NASA/IPAC Extragalactic Database (NED), which is operated
by the Jet Propulsion Laboratory, California Institute of
Technology, under contract with the National Aeronautics and Space
Administration ({\tt http://nedwww.ipac.caltech.edu/}); of the
data from the Two Micron All Sky Survey (2MASS, {\tt
http://www.ipac.caltech.edu/2mass/} \linebreak {\tt
releases/allsky/}), which is a joint project of the University of
Massachusetts and the Infrared Processing and Analysis
Center/California Institute of Technology, funded by the National
Aeronautics and Space Administration and the National Science
Foundation, and the Sloan Digital Sky Survey (SDSS). Funding for
the SDSS and SDSS-II has been provided by the Alfred P. Sloan
Foundation, the Participating Institutions, the National Science
Foundation, the U.S. Department of Energy, the National
Aeronautics and Space Administration, the Japanese Monbukagakusho,
the Max Planck Society, and the Higher Education Funding Council
for England. The SDSS Web Site is {\tt http://www.sdss.org/}.

The SDSS is managed by the Astrophysical Research Consortium for the Participating Institutions. The Participating
Institutions are the American Museum of Natural History, Astrophysical Institute Potsdam, University of Basel,
University of Cambridge, Case Western Reserve University, University of Chicago, Drexel University, Fermilab, the
Institute for Advanced Study, the Japan Participation Group, Johns Hopkins University, the Joint Institute for Nuclear
Astrophysics, the Kavli Institute for Particle Astrophysics and Cosmology, the Korean Scientist Group, the Chinese
Academy of Sciences (LAMOST), Los Alamos National Laboratory, the Max-Planck-Institute for Astronomy (MPIA), the Max-
Planck-Institute for Astrophysics (MPA), New Mexico State University, Ohio State University, University of Pittsburgh,
University of Portsmouth, Princeton University, the United States Naval Observatory, and the University of Washington.
\end{acknowledgments}

\newpage
\begin{center}
\refname
\end{center}


\begin{thebibliography}{99}
\bibitem{Hayashi:Kopylov_n}
E.~Hayashi, S.\,D.\,M.~White, \mnras~ {\bf 370}, L38 (2006).
\bibitem{Kop1:Kopylov_n}
A.~I.~Kopylov and F.~G.~Kopylova, \ab~ {\bf 62}, 311 (2007).
\bibitem{Chincarini:Kopylov_n}
G.~Chincarini, H.~J.~Rood, G.~N,~Sastry, and G.~A.~Welch, \apj~ {\bf 168}, 11 (1971).
\bibitem{Miley:Kopylov_n}
G.~K.~Miley and D.~E.~Harris, \aaa~ {\bf 61}, L23 (1977).
\bibitem{Owen:Kopylov_n}
F.~N.~Owen and M.~J.~Lewlow, \apjs~ {\bf 108}, 41 (1997).
\bibitem{Bliton:Kopylov_n}
M.~Bliton, E.~Rizza, J.~O.~Burns, et al., \mnras~ {\bf 301}, 609 (1998).
\bibitem{Giovan:Kopylov_n}
G.~Giovannini and L.~Feretti, New Astron. {\bf 5}, 335 (2000).
\bibitem{Zabludoff:Kopylov_n}
A.~I.~Zabludoff, J.~P.~Huchra, and M.~J.~Geller, \apjs~ {\bf 71}, 1 (1990).
\bibitem{Oegerle:Kopylov_n}
W.~R.~Oegerle, J.~M.~Hill, and M.~J.~Fitchett, \aj~ {\bf 110}, 32 (1995).
\bibitem{Girardi:Kopylov_n}
M.~Girardi, G.~Giurisin, F.~Mardirossian, et al.,  \apj~ {\bf 505}, 74 (1998).
\bibitem{Jones:Kopylov_n}
C.~Jones and W.~Forman, \apj~ {\bf 276}, 38 (1984).
\bibitem{Vikhlinin:Kopylov_n}
A.~Vikhlinin, B.~R.~ McNamara, W.~Forman, et al.,  \apj~ {\bf 502}, 558 (1998).
\bibitem{Reiprich:Kopylov_n}
T.~H.~Reiprich, H.~ B\"{o}hringer, \apj~ {\bf 567}, 716 (2002).
\bibitem{Abell:Kopylov_n}
G.~O.~Abell, H.~G.~Jr.~Corwin, and R.~P.~Olowin, \apjs~ {\bf 70}, 1 (1989).
\bibitem{Adelman:Kopylov_n}
J.~K.~Adelman-McCarthy et al., \apjs~ {\bf 175}, 297 (2008).
\bibitem{Kop2:Kopylov_n}
F.~G.~Kopylova and A.~I.~Kopylov, \ab~ {\bf 64}, 1 (2009).
\bibitem{Anders:Kopylov_n}
K.~Andersson, J.~R.~Peterson, G.~Madejski, and A.~Goober, astro-ph/0902.0003.
\bibitem{Landolt:Kopylov_n}
A.~U.~Landolt, \aj~ {\bf 104}, 340 (1994).
\bibitem{Sersic:Kopylov_n}
J.~L.~S\'{e}rsic, Bol. Asoc. Argent. Astron. {\bf 6}, 41 (1963).
\bibitem{deV:Kopylov_n}
G.~de~Vaucouleurs, Ann. d'Astrophys. {\bf 11}, 247 (1948).
\bibitem{Saglia:Kopylov_n}
R.~P.~Saglia, E.~Bertschinger, G.~Baggley, et al., \mnras~ {\bf 264}, 961 (1993).
\bibitem{Bern3:Kopylov_n}
M.~Bernardi, R.~K.~Sheth, R.~Nichol, et al., \mbox{\aj}~ {\bf 129}, 61 (2005).
\bibitem{Lupton:Kopylov_n}
R.~Lupton, SDSS DR(4), (2005).
\bibitem{Kormendy:Kopylov_n}
J.~Kormendy, \apj~ {\bf 218}, 333 (1977).
\bibitem{Graham:Kopylov_n}
A.~W.~Graham, \mnras~ {\bf 334}, 859 (2002).
\bibitem{Djorgovski:Kopylov_n}
S.~Djorgovski and M.~Davis, \apj~ {\bf 313}, 59 (1987).
\bibitem{Kop3:Kopylov_n}
F.~G.~Kopylova and A.~I.~Kopylov, Astron. Lett. {\bf 27}, 345 (2001).
\bibitem{Kop4:Kopylov_n}
A.~I.~Kopylov and F.~G.~Kopylova, \aaa~ {\bf 382}, 389 (2002).
\bibitem{Bern1:Kopylov_n}
M.~Bernardi, R.~K.~Sheth, J.~Annis, et al., \mbox{\aj}~ {\bf 125}, 1817 (2003a).
\bibitem{Bern2:Kopylov_n}
M.~Bernardi, R.~K.~Sheth, J.~Annis, et al., \mbox{\aj}~ {\bf 125}, 1866 (2003b).
\bibitem{Kop5:Kopylov_n}
F.~G.~Kopylova and A.~I.~Kopylov, Astron. Lett. {\bf 33}, 211 (2007).
\end{thebibliography}
\end{document}